\documentclass[american]{llncs}
\usepackage[nodate]{datetime}
\usepackage{babel}
\usepackage{amsmath}
\usepackage{mathtools}[2011/02/12]
\usepackage{amsfonts,amssymb,stmaryrd}

\usepackage{MnSymbol}
\usepackage[fancyproofs,noextended,squareitemtag]{theorems}[2013/01/24]

\newcommand*{\dfn}{\coloneq}
\newcommand*{\seq}[2][n]{#2_{1}, \dots, #2_{#1}}
\providecommand*{\mathoper}[1]{\mathop{\mathit{#1}}\nolimits}
\providecommand*{\pair}[2]{{\langle #1, \, #2 \rangle}}
\providecommand*{\triple}[3]{{\langle #1, \, #2, \, #3 \rangle}}
 
\providecommand*{\ie}   {i.e.,} 
\providecommand*{\resp} {respectively}

\providecommand*{\wrt}  {w.r.t.}
\makeatletter
\providecommand{\ifempty}[3]{\def\@@@temp{#1}\ifx\@@@temp\@empty#2\else#3\fi}
\makeatother
\providecommand*{\wgmWarn}[1]{\typeout{*** using WgMacros def of \string#1}}

\providecommand*{\parensmathoper}[2]{\ensuremath{\mathoper{#1}\ifempty{#2}{}{(#2)}}}

\providecommand*{\newsmartprefix}[2]{\wgmWarn{\newsmartprefix}}
\providecommand*{\newsmartsprefix}[2]{\wgmWarn{\newsmartsprefix}}
\providecommand*{\newdefinition}[1]{\wgmWarn{\newdefinition}\newdefinitionaux}
\providecommand*{\newdefinitionaux}[2][]{}
\newcommand{\ands}{\fbox{this should not be used, is just to prevent 
    unobserved redefinition of command \string\ands}}
\newcommand*{\sequent}[4][]{\gdef\and{\quad\hfill}
    \gdef\ands{\;\hfill\ldots\hfill\;}\global\setlabel{#1}
    {#1}\frac{\:\:#2\:\:}{\;#3\;}\;\text{\small%
\begin{tabular}{@{}l@{}} #4 \end{tabular}}}

\newcommand*{\R}{\mathbb{R}}
\newcommand*{\Conf}{\mathit{Conf}}
\newcommand*{\clauseif}{\ensuremath{\mathrel{\mathord{:}\mathord{-}}}}
\newcommand*{\ra}{\rightarrow}
\newcommand*{\nra}{\not\rightarrow}
\newcommand*{\tccp}{\textit{tccp}}
\newcommand*{\ccp}{\textit{ccp}}

\newcommand*{\hcc}{\textit{hcc}}
\newcommand*{\hytccp}{{\textsc hy}-\-\tccp}
\newcommand*{\askip}{\mathord{\mathsf{stop}}}
\newcommand*{\atell}{\parensmathoper{\mathsf{tell}}}
\newcommand*{\aask}[1]{\mathop{\mathsf{ask}}\ifempty{#1}{}{(#1)\ra}}
\newcommand*{\asumask}[4][i]{\sum_{#1=1}^{#2}\aask{#3_{#1}}{#4}_{#1}}
\newcommand*{\anow}[3]{\tagnow\ifempty{#1}{} {\: #1 \tagthen #2\ifempty{#3}{}{\tagelse #3}}}

\newcommand*{\ahiding}[3][]{\mathop{\exists^{#1}{#2}} {#3}}

\newcommand*{\tagelse}{\mathrel{\mathsf{else}}}
\newcommand*{\tagnow}{\mathop{\mathsf{now}}}
\newcommand*{\tagthen}{\mathrel{\mathsf{then}}}
\newcommand*{\aaskc}[1]{\tagaskc \ifempty{#1}{}{(#1)}}
\newcommand*{\achange}[3]{\tagchange \ifempty{#1}{}{(#1,#2,#3)}}
\newcommand*{\tagchange}{\mathop{\mathsf{change}}}
\newcommand*{\tagaskc}{\mathop{\widetilde{\mathsf{ask}}}}

\newcommand*{\CSdom}{\mathcal{C}} 
\newcommand*{\CSfalse}{\mathit{false}}
\newcommand*{\CShid}[2]{\mathop{\exists}\nolimits_{#1} #2}
\newcommand*{\CSimp}{\mathrel{\vdash}}
\newcommand*{\CSnimp}{\mathrel{\nvdash}}

\newcommand*{\CSmerge}{\mathbin{\wedge}}

\newcommand*{\CStrue}{\mathit{true}}
\newcommand*{\CSys}{\mathbf{C}}
\newcommand*{\Var}{\mathit{Var}}
\newcommand{\Time}{T}
\newcommand{\Inflows}{\mathit{In}}
\newcommand{\Outflow}{\mathit{Out}}
\newcommand{\Vol}{\mathit{Vol}}
\newcommand{\ToG}{\mathit{ToG}}
\newcommand{\agate}{\mathtt{gate}}
\newcommand{\close}{\mathit{close}}
\newcommand{\half}{\mathit{halfOpen}}
\newcommand{\all}{\mathit{open}}
\newcommand{\INITVOL}{\mathit{INITVOL}}
\newcommand{\THRESHOLD}{\mathit{THRESHOLD}}
\newcommand{\NewIn}{\mathit{NewIn}}
\newcommand{\acontrol}{\texttt{controller}}
\newcommand{\asupplier}{\mathtt{supplier}}
\newcommand*{\HVar}{\widetilde{\mathit{Var}}}
\newcommand*{\HVarTuple}[3]{#1\mapsto(#2,#3)}
\newcommand*{\HVarUpdate}[2]{\ifempty{#1}{\lhd}{#1 \lhd #2}}
\newcommand*{\inv}{\mathit{inv}}
\newcommand*{\dt}{\sigma}
\newcommand*{\rad}{\ra_{\dt}}
\newcommand*{\rac}[1]{\ra_{\mathit{#1}}}

\newcommand*{\hst}[2]{\langle #1,\, #2 \rangle}
\newcommand*{\Confc}{\widetilde{\mathit{Conf}}}

\begin{document}


\title{Modeling Hybrid Systems in \textsc{Hy}-\tccp}
\author{
Dami\'an Adalid \qquad Mar\'ia del Mar Gallardo \qquad Laura Titolo
\institute{Dept. Lenguajes y Ciencias de la Computaci\'on\\
E.T.S.I. Inform\'atica \quad
University of M\'alaga\thanks{This work has been
  supported by the Andalusian Excellence Project P11-TIC7659 and the Spanish Ministry of  Economy and Competitiveness project TIN2012-35669 }}
\email{[damian,gallardo,laura.titolo]@lcc.uma.es }
}

\maketitle

\def\authorrunning{Adalid, Gallardo \& Titolo}
\def\titlerunning{\textsc{Hy}-\tccp: a hybrid extension of \tccp}

\begin{abstract}
    Concurrent, reactive and hybrid systems
    require quality modeling languages to be described and analyzed.
    The Timed Concurrent Constraint Language (\tccp)
    was introduced as a simple but powerful model
    for reactive systems.
    In this paper, we present \emph{hybrid \tccp} (\hytccp),
    an extension of \tccp{} over continuous time which
    includes new constructs to model the continuous dynamics of hybrid systems.
\end{abstract}

\section{Introduction}
\label{sec:intro}
Concurrent, reactive and hybrid systems 
have had a wide diffusion and they have become essential
to an increasingly large number of applications.
Often, systems of these kinds are safety critical,
\ie{} an error in the software can have
tragic consequences.
In the case of hybrid systems, the modeling and the analysis
phases are particularly hard due to the combination of discrete 
and continuous dynamics and the presence of real variables.
Many formalisms have been developed to describe concurrent systems.
One of these is the \emph{Concurrent Constraint paradigm} (\ccp)
\cite{Saraswat89phd}. It differs from other paradigms mainly due to the
notion of store-as-constraint that replaces the classical store-as-valuation model.
In this paradigm, the agents running in parallel
communicate by means of a global constraint store.
The \emph{Timed Concurrent Constraint Language}
\cite{deBoerGM99} (\tccp{} in short) is a concurrent logic language obtained
by extending \textit{ccp} with the notion of time and a suitable
mechanism to model time-\-outs and preemptions.

In this paper, we present \hytccp{} an extension of \tccp{} over continuous time.
The declarative nature of \hytccp{} facilitates
a high level description of hybrid systems
in the style of hybrid automata \cite{Henzinger96}.
Furthermore, its logical nature
eases the development of semantics
based program manipulation tools
for hybrid systems
(verifiers, analyzers, debuggers\dots).
Parallel composition of hybrid automata is naturally
supported in \hytccp{} due to the existence of a global shared store and to
the synchronization mechanism.

The paper is organized as follows. In Section~\ref{sec:extension},
we briefly introduce the language \tccp{}
and we show how we have extended it to obtain \hytccp.
Section~\ref{sec:examples} contains an example to highlight the expressive power of our language. Finally,
Section~\ref{sec:conclusions} concludes the paper and presents some related work.

\section{\textsc{Hy}-\tccp: a hybrid extension of \tccp}
\label{sec:extension}
The \emph{Timed Concurrent Constraint Language} (\tccp, \cite{deBoerGM99})
is a time extension of \ccp{} suitable for describing
concurrent and reactive systems.
The computation in  \tccp{}
proceeds as the concurrent execution
of several agents that can monotonically add information in
a global constraint \emph{store}, or query information from it.
\tccp{} is parametric
\wrt\ a \emph{cylindric
constraint system} which handles the information on system variables.
Briefly, a cylindric
constraint system is 
a structure $\CSys=
\langle \CSdom, \mathord{\CSimp}, \CSmerge, \CSfalse, \CStrue,
\mathord{\Var}, \CShid{}{} \rangle$ composed by a set of constraints
$\CSdom$
ordered by the entailment relation $\CSimp$ (intuitively, $c\CSimp d$
if $c$ contains more information than $d$)
where $\CSmerge$ is a binary operator
that merges the information from two constraints;
$\CSfalse$ and $\CStrue$ are,
respectively, the greatest and the least element of $\CSdom$; $\Var$ is a
denumerable set of variables and $\CShid{}$ existentially quantifies
variables over constraints.
The syntax of agents is given by the grammar:
\begin{align*}
    A ::= \askip \mid \atell{c} \mid A \parallel A \mid
    \ahiding{x}{A} \mid
    \textstyle{\sum_{i=1}^{n}\aask{c_{i}} A}
    \mid \anow{c}{A}{A}
    \mid p(\bar{x})
\end{align*}
where $c$, $\seq{c}$ are finite constraints in $\CSdom$,
$\bar{x} \in \Var\times\dots\times\Var$
and $p$ is a predicate symbol.
A \tccp{} program
is a pair $D.A$, where $A$ is the initial agent
and $D$ is a set of \emph{process declarations} of
the form $p(\bar{x}):- A$.  The notion of time is
introduced by defining a \emph{discrete} global clock.
The \emph{operational semantics} of \tccp{} \cite{deBoerGM99}
is described by a transition system $T=(\Conf, \ra)$.
Configurations in $\Conf$ are pairs $\pair{A}{c}$ representing the agent
$A$ to be executed in the current global store $c$.  The transition
relation ${\ra} \subseteq \Conf\times\Conf$ is the least relation
satisfying the rules in Figure~\ref{fig:op_sem_tccp}.
As can be seen from the rules, the $\askip$ agent represents the successful
termination of the computation.
The $\atell{c}$ agent adds the constraint $c$ to the current store.
The choice agent $\asumask{n}{c}{A}$
non-deterministically executes
one of the
agents $A_i$ whose corresponding guard $c_i$ is entailed by the 
store; otherwise, if no guard is entailed by the store, the agent suspends.
The conditional agent $\anow{c}{A}{B}$ behaves
like $A$ (\resp\ $B$) if $c$ is (\resp\ is not) entailed by the store.
$A\parallel B$ models the parallel composition of $A$ and $B$ in terms of
maximal parallelism,
\ie\ all the enabled agents of $A$ and $B$ are executed at the same time.
The agent $\ahiding{x}{A}$ makes variable $x$ local to $A$.
Finally, the agent $p(\bar{x})$
takes from $D$ a declaration of the form
$p(\bar{x}):-{A}$ and then executes $A$.

\begin{figure}[t]
    \begin{minipage}{\textwidth}
        {\scriptsize
        {\setlength{\jot}{1ex}
        \begin{align*}
            & \sequent{}
            {\pair{\atell{c}}{d} \ra \pair{\askip}{c \CSmerge d}
            }
            {}{}
            & \sequent{\exists\,1 \leq j \leq n,\, d \CSimp c_{j} }
            {\pair{\asumask{n}{c}{A}}{d} \ra \pair{A_{j}}{d} }
            {}{}
            \\
            & \sequent{\pair{A}{d} \ra \pair{A'}{d'},\, d\CSimp c }
            {\pair{\anow{c}{A}{B}}{d} \ra \pair{A'}{d'} }
            {}{}
            & \sequent{ \pair{A}{d} \nra,\, d\CSimp c }
            { \pair{\anow{c}{A}{B}}{d} \ra
            \pair{A}{d} }
            {}{}
            \\
            & \sequent{ \pair{B}{d} \ra \pair{B'}{d'},\, d\CSnimp c }{
            \pair{\anow{c}{A}{B}}{d} \ra \pair{B'}{d'} }
            {}{}
            & \sequent{ \pair{B}{d} \nra ,\, d\CSnimp c  }{ \pair{\anow{c}{A}{B}}{d} \ra
            \pair{B}{d} }
            {}{}  
            \\
            & \sequent{ \pair{A}{d} \ra \pair{A'}{d'} \quad \pair{B}{d} \ra
            \pair{B'}{c'} }{ \pair{A\parallel B}{d} \ra \pair{A'\parallel
            B'}{d'\CSmerge c'} }
            { }{}
            & \sequent{ \pair{A}{d} \ra \pair{A'}{d'} \quad \pair{B}{d} \nra }{
                \pair{A\parallel B}{d} \ra \pair{A'\parallel B}{d'} 
            }{ }{}
            \\
            & \sequent{ \pair{A}{l \CSmerge \CShid{x}{d}} \ra\pair{B}{l'} }{
            \pair{\ahiding[l]{x}{A}}{d} \ra \pair{\ahiding[l']{x}{B}}{d
            \CSmerge \CShid{x}{l'}} }{ }{}
            & \sequent{ p(\bar{x})\clauseif A \in D }{ \pair{ p(\bar{x}) }{d} \ra \pair{A}{d}
            }{}{}
        \end{align*}
        }
        }
        \caption[The transition system for \tccp{}.]{The transition system for \tccp{}.\footnotemark{}}
        \label{fig:op_sem_tccp}
    \end{minipage}
\end{figure}

\footnotetext{The auxiliary agent $\ahiding[l]{x}{A}$ makes explicit the
local store $l$ of $A$.  This auxiliary agent is linked to the principal
hiding construct by setting the initial local store to $\CStrue$, thus
$\ahiding{x}{A} \dfn \ahiding[\CStrue]{x}{A}$.}

We introduce the language \hytccp,
which subsumes \tccp{} and includes new agents to model
hybrid systems in the style of hybrid automata.
Hybrid automata \cite{Henzinger96} are an extension of finite-state automata.
Intuitively, their discrete behavior is defined by means of a finite set of discrete states
(called {\em locations}) and a set of (instantaneous) \emph{discrete transitions} from one location to another.
The continuous behavior of hybrid automata
is described at each location by some
Ordinary Differential Equations (ODEs) which describe
how continuous variables evolve over time
(\emph{continuous transitions}).
Each location is associated with an \emph{invariant} predicate which constrains the value of the
continuous variables at that location and
with an \emph{initial} predicate that establishes their possible initial values.
Discrete transitions are associated with a \emph{jump} predicate that may include a \emph{guard}
and a \emph{reset} predicate which
updates the value and/or the flow of continuous variables.

\hytccp{} uses a \tccp{} monotonic store (called \emph{discrete store}) to model the
information about the current location and the associated
invariants.
Discrete transitions of hybrid automata are
modeled as instantaneous
transitions in \hytccp{} and they are used to synchronize parallel agents.
We distinguish the set of discrete variables $\Var$, whose
information is accumulated monotonically,
and the set of continuous variables $\HVar$,
whose values change continuously over time ($\Var\cap\HVar = \emptyset$).
Constraints in $\CSdom$ are now defined over $\Var\cup\HVar$.
The \tccp{} store is extended by adding a
component called \emph{continuous store}.
The continuous store is not monotonic, instead it records the dynamical evolution of the continuous
variables.
Thus, a \hytccp{} store is a pair $\hst{c}{\tilde{c}}$ where
$c$ (discrete store) is a monotonic constraint store as in \tccp{} and
$\tilde{c}$ (continuous store) is a function that associates
a continuous variable with its current value and its flow\footnote{In this paper,
we assume that continuous variables evolve independently
from each other.
Given $x\in\HVar$, its flow is defined as a predicate on set $\{x,\dot{x}\}$
where $\dot{x}$ denotes the first order derivative of $x$
(e.g. $\dot{x} = 2$ or $\dot{x} = 2x$).
},
which indicates how its value changes over time by means of an ODE.
Given a continuous store $\tilde{c}$ and a continuous variable $x$,
$\tilde{c}(x) = \pair{v}{f}$ means that $x$ has value $v$ and flow $f$.
Given $\tau\in\mathbb R_{> 0}$ and
$\inv\in\CSdom$ we denote as $\hst{c}{\tilde{c}} \rightsquigarrow_\tau^{\inv} \hst{c}{\tilde{c}_\tau}$ the
projection of the store $\hst{c}{\tilde{c}}$ at time $\tau$ satisfying $\inv$.
The value of the variables are
updated at time $\tau$, while the flows are unchanged.
In order to model behaviors typical of hybrid
systems we introduce two new agents \wrt{} \tccp{}: $\tagchange$ and $\tagaskc$.
The agent $\tagchange$ updates the value and/or the flow of a given continuous variable
(\emph{reset} predicate of hybrid automata).
The \tccp{} choice agent is extended by allowing the non-deterministic
choice between discrete and continuous transitions in the following way:
$\textstyle{\sum_{i=1}^{n}\aask{c_{i}} A_i + \sum_{j=1}^{m} \aaskc{\inv_{j}}}$
where $n\geq 0$ and $m\geq 0$.
Here, the $\tagaskc$ branches can be non-\-deterministically selected
in case the invariant $inv_j$ is entailed in the current store. This corresponds
to the passage of continuous time in a hybrid automaton location.
The continuous variables evolve over \emph{continuous} time while $inv_j$ holds and
until another ask branch is selected.
The \emph{operational semantics} of \hytccp{}
is described by a transition system $T=(\Confc, \rad, \rac{\tau})$.
Configurations in $\Confc$ are triple $\triple{A}{c}{\tilde{c}}$
representing the agent
$A$ to be executed in the current extended store $\hst{c}{\tilde{c}}$.
The transition relation $\rad \subseteq \Confc\times\Confc$
represents a \tccp{} discrete transition whose execution is
instantaneous, while
$\rac{\tau} \subseteq \Confc\times\Confc$ models a
continuous transition of duration $\tau$.
In Figure~\ref{fig:op_sem_hytccp} we describe
the rules that we have added to the operational semantics of \tccp{}
in order to deal with continuous time and variables.
In Rule~\ref{rule:R1} the agent $\tagchange$
uses the operator $\HVarUpdate{}{}$ 
that, given a continuous store $\tilde{c}$
and a triple $(x,v,f)$, updates $\tilde{c}$ with
a new initial value $v$ and a new flow $f$ for the variable $x$.
In Rule~\ref{rule:R2}
time passes continuously while one of the $\tagaskc$ invariants holds in the store and 
the values of the continuous variables change over time following their flow.
Rule~\ref{rule:R3} represents the parallel execution of
two continuous transitions, note that their duration must coincide.
Rule~\ref{rule:R4} expresses the parallel composition of a discrete and a continuous transition.
In this case, the discrete transition is executed before the continuous one.

\begin{figure}[t]
    \begin{minipage}{\textwidth}
        {\scriptsize
        {\setlength{\jot}{1ex}
        \begin{align*}
            &\sequent{}
                {
                \triple{\achange{x}{v}{f}}{d}{\tilde{d}}
                \rad
                \triple{\askip}{d}{\HVarUpdate{\tilde{d}}{(\HVarTuple{x}{v}{f})}}
                }{}{\tag{\textbf{R1}}}\label{rule:R1} 
            \\            
            & \sequent{\exists\,1 \leq k \leq m,\,\tau\in\R^+.\, \hst{d}{\tilde{d}} \rightsquigarrow_\tau^{inv_j}\hst{d}{\tilde{d}_{\tau}}}
            {\triple{\textstyle{\sum_{i=1}^{n}\aask{c_{i}} A_i + \sum_{j=1}^{m}\aaskc{\inv_j}}}{d}{\tilde{d}}
             \rac{\tau}
             \triple{\textstyle{\sum_{i=1}^{n}\aask{c_{i}} A_i + \sum_{j=1}^{m}\aaskc{\inv_j}}}{d}{\tilde{d}_\tau}
             }
            {}{\tag{\textbf{R2}}}\label{rule:R2}
            \\
            & \sequent{ \triple{A}{d}{\tilde{d}} \rac{\tau} \triple{A}{d}{\tilde{d}'} \quad
            \triple{B}{d}{\tilde{d}} \rac{\tau} \triple{B}{d}{\tilde{d}'} }{
                \triple{A \parallel B}{d}{\tilde{d}} \rac{\tau} \triple{A \parallel B}{d}{\tilde{d}'} 
            }{}{\tag{\textbf{R3}}}\label{rule:R3} 
           \\
           & \sequent{ \triple{A}{d}{\tilde{d}} \rad \triple{A'}{d'}{\tilde{d}'} \quad
           \triple{B}{d}{\tilde{d}} \rac{\tau} \triple{B}{d}{\tilde{d}''} }{
               \triple{A \parallel B}{d}{\tilde{d}} \rad \triple{A'\parallel B}{d'}{\tilde{d}'}
           }{}{\tag{\textbf{R4}}}\label{rule:R4}       
        \end{align*}
        }
        }
        \caption[A fragment of the transition system for \hytccp{}.]{The transition system for \hytccp{}.}
        \label{fig:op_sem_hytccp}
    \end{minipage}
\end{figure}

\section{Example: a dam management system}
\label{sec:examples}
In this section we model a dam management system with \hytccp{} (Figure~\ref{fig:tccp model}).
Our experience in this area \cite{GallardoMPL11}
has shown us that this is a
realistic and significant example to demonstrate the expressive power and usability of our language.
Due to the monotonicity of the discrete constraint store, streams (written in a
list-fashion way) are used to model \emph{imperative-style} variables
\cite{deBoerGM99}.
Our dam controller system is modeled
as the parallel composition of a controller,
a supplier and two gate processes\footnote{The code of $\agate$ is
omitted due to space limitations.}.
$\Vol$ represents the total amount of water,
it has initial value $\INITVOL$ and flow $0$ (\ie{} its value is constant over time).
$\Time$ represents a timer used by the \texttt{supplier},
it has initial value $0$ and flow $1$,
thus it evolves lineary over time. 
When $\Time$ reaches
the value 3600, the \texttt{supplier} sends
to the \texttt{controller}
the value of the new inflow of water through the input
channel $\Inflows$.
At this point, the \texttt{controller} checks to which interval
the current volume of water ($\Vol$) belongs. Intervals are defined by using
several sub-\-indexed constants $\THRESHOLD_i$.
According to the current value of $\Vol$,
the \texttt{controller} sends a signal
to each gate through the output channels $\ToG1$ and $\ToG2$
in order to set their status.
At the same time, the continuous store is updated.
We use the symbol $\_$ in the second argument of agent $\tagchange$
to indicate that only the flow of $\Vol$ is updated, while its value is unchanged.
The new flow of $\Vol$ depends on the new inflow
received from the \texttt{supplier} ($\NewIn$) and on a value representing the
water discharged through the gates.
This value is computed by the function $\Outflow$ according to
the current state of the gates.
It is worth noting that the $\tagaskc$ construct in \texttt{supplier} is used
to make time pass. Its invariant
ensures that $\Time$ never exceeds the value $3600$.

\begin{figure}[ht!]
{\scriptsize
\begin{center}
\begin{tabbing}
\texttt{init}\=$\clauseif \exists\ \Inflows,\, \ToG1, \ToG2, \Vol, \Time
  \big(\achange{\mathit{\Time}}{0}{\dot{\Time} {=} 1}\parallel \asupplier (\Time,\,\Inflows) \parallel \atell{\Vol{\leq}\THRESHOLD_3} \parallel$\\
$\quad\achange{\Vol}{\INITVOL}{\dot{\Vol}{=}0} \parallel \acontrol(\Vol, \Inflows, \ToG1, \ToG2)  \parallel \agate(\mathit{ToG}1) \parallel \agate(\mathit{ToG2})\big)$ \\[1.5ex]
$\acontrol (\Vol, \Inflows, \ToG1, \ToG2) \clauseif \exists \NewIn, \ToG1', \ToG2', \Inflows'\Big( $\\
\=
$\quad\aask{\Inflows{=}[\NewIn|\_]} \big(\atell{\Inflows{=}[\NewIn|\Inflows']} \parallel$ \\ 
\>\qquad\=$\aask{\Vol{\leq}\THRESHOLD_1} (\atell{\ToG1{=}[\close|\ToG1']} \parallel\atell{ToG2{=}[\close|ToG2']} \parallel $ \\
\>\>$\qquad\achange{\Vol}{\mathit{\_}}{\dot{\Vol}{=}\NewIn{-}\Outflow(\close,\close)} \parallel \acontrol(\Vol, \Inflows',\ToG1', \ToG2'))$ \\
\>$\qquad + \aask{\Vol {>}\THRESHOLD_1 \wedge \Vol{\leq}\THRESHOLD_2}(\atell{ToG1{=}[\half|ToG1']} \parallel$\\
\>\>  $\qquad\atell{\ToG2{=}[\half|\ToG2']} \parallel \achange{\Vol}{\mathit{\_}}{\dot{\Vol} {=} \NewIn {-} \Outflow(\half,\half)} \parallel $\\
\>\>  $\qquad\acontrol(\Vol, \Inflows',\ToG1',\ToG2'))$\\
\> $\qquad+ \aask{\Vol{>}\THRESHOLD_2 \wedge \Vol{<}\THRESHOLD_3}(\atell{\ToG1{=}[\half|\ToG1']} \parallel$\\
\>\> $\qquad\atell{\ToG2{=}[\all|\ToG2']} \parallel \achange {\Vol}{\mathit{\_}} {\dot{\Vol} {=} \NewIn{-}\Outflow(\half, \all)} \parallel $ \\
\>\> $\qquad\acontrol(\Vol, \Inflows',\ToG1',\ToG2'))\big)$\\
$\quad + \aask{\Vol{=}\THRESHOLD_3}\big(\atell{\ToG1{=}[\all|\ToG1']} \parallel \atell{\ToG2{=}[\all|\ToG2']} \parallel $ \\
\>\> $\qquad\achange {\Vol}{\mathit{\_}} {\dot{\Vol} {=} {-}\Outflow(\all, \all)} \parallel \acontrol(\Vol, \Inflows,\ToG1',\ToG2')\big)\Big)$ \\[1.5ex]
\=$\asupplier(\Time,\, \Inflows) \clauseif \exists \Inflows' \big( \aaskc{\Time{\leq} 3600}$ \\
\>$\quad+ \aask{\Time{=}3600}$
$(\atell{\Inflows{=}[\mathit{Random}(0,350)|\Inflows']}
\parallel \achange{T}{0}{\dot{T} {=} 1} \parallel \asupplier(\Time,\,\Inflows'))\big)$\\
\end{tabbing}
\end{center}
\caption{{\sc hy}-\tccp\  model for a dam management system}
\label{fig:tccp model}
}
\end{figure}

\section{Conclusions and Related Work}
\label{sec:conclusions}

In this paper we have presented \hytccp{}, an extension of \tccp{}
over continuous time,
with the aim of
modeling hybrid systems in a simple and declarative way.
The language is parametric to both, the cylindric constraint system used to
manage the discrete behavior, and the class of differential equation solvers
that models the continuous behavior.
Although we are aware that the decidability limits of hybrid systems~\cite{Henzinger96}
lie on the class of initialized rectangular systems,
in this paper we have only restricted the class of differential equations used by
assuming that the dynamics of a continuous variable does not depend
on the others. In this way we obtain a more general framework.

In \cite{GuptaJSB94}, \hcc{} was introduced
as the first extension over continuous time of the concurrent
constraint paradigm.
Although both \hytccp{} and \hcc{} are declarative languages with a logical nature,
they have some important differences.
\hytccp{} has been defined as a  modeling language for hybrid systems in the
style of hybrid automata.
Unlike \hcc{}, which is deterministic,
\hytccp{} provides the non-deterministic choice agent which allows the transitions of 
hybrid automata to be expressed as a list of $\aask{}$ and $\aaskc{}$ branches.
Furthermore, in \hcc{}, the information on the value and flow of continuous variables is modeled
as a constraint of the underlying continuous constraint system.
On the contrary, in \hytccp{}, there is a clear distinction between discrete and continuous variables.
The process algebra \textit{Hybrid Chi} \cite{BeekMRRS06} shares with \hytccp{}
the separation between
discrete and continuous variables, the synchronous nature and
the concept of delayable guard (corresponding to the suspension of the non-deterministic choice).
In \cite{CuijpersR05}, \textit{HyPa} is introduced as
an extension of the process algebra \textit{ACP}.
It differs from \textit{Hybrid Chi}
mainly in the way time-determinism is treated, and in the modeling of time passing.

In the future we plan to
develop a framework for the description and simulation of
\hytccp{} programs.
We are also interested in defining
a translation rules system from \hytccp{} to hybrid automata and viceversa.
Furthermore, we plan to use model checking and abstract interpretation to
verify temporal properties of hybrid systems written in \hytccp{} (as done in \cite{GallardoP13} for SPIN).

\bibliographystyle{splncs03}
\bibliography{biblio}
\newcommand{\etalchar}[1]{$^{#1}$}

\end{document}